\documentclass[onecolumn,aps,prd,preprintnumbers,superscriptaddress,nofootinbib,amsmath,amssymb,floats,floatfix,showkeys,notitlepage,showpacs]{revtex4-2}

\usepackage{comment}
\usepackage{tabularx}
\usepackage{lipsum}
\usepackage{graphicx}
\usepackage{subfigure}
\usepackage{palatino}
\usepackage{sans}
\usepackage{hyperref}
\hypersetup{colorlinks=true,linkcolor=blue,urlcolor=blue,citecolor=blue}
\usepackage[toc,page]{appendix}
\usepackage[normalem]{ulem}
\usepackage{adjustbox}
\usepackage{latexsym}
\usepackage{amsmath}
\usepackage{amssymb}
\usepackage{amsfonts}
\usepackage{dcolumn}
\usepackage{bm}
\usepackage{tikz}
\usepackage{bigints}
\usepackage{array,tabularx,multirow,booktabs}
\usepackage[tracking=true]{microtype}
\SetTracking{}{500}
\SetTracking{encoding={*}, shape=sc}{40}
\UseRawInputEncoding %for inputenc error%
\allowdisplaybreaks
\begin{document} \sloppy
\title{Analyzing the effect of higher dimensions on the black hole silhouette, deflection angles, and PINN approximated quasinormal modes}

\author{Nikko John Leo S. Lobos}
\email{nslobos@ust.edu.ph}
\affiliation{Electronics Engineering Department, University of Santo Tomas, Espa\~{n}a Blvd, Sampaloc, Manila, 1008 Metro Manila}
\author{Anele M. Ncube}%
\email{ancube@uj.ac.za}
\affiliation{Department of Physics, University of Johannesburg, PO Box 524, Auckland Park 2006, South Africa}
\author{Reggie C. Pantig}
\email{rcpantig@mapua.edu.ph}
\affiliation{Physics Department, Map\'ua University, 658 Muralla St., Intramuros, Manila 1002, Philippines}
\author{Alan S. Cornell}
\email{acornell@uj.ac.za}
\affiliation{Department of Physics, University of Johannesburg, PO Box 524, Auckland Park 2006, South Africa}

\begin{abstract}
This study investigates the effects of higher dimensions on the observable properties of Schwarzschild-Tangherlini black holes, focusing on the photonsphere, shadow radius, deflection angles, and quasinormal modes (QNMs). By extending classical methods with Physics-Informed Neural Networks (PINNs), the research examines how increasing dimensionality alters these properties, causing shadow size reduction, weaker deflection angles, and shifts in QNM frequencies. The findings suggest that as black holes increase in dimensionality, their gravitational influence diminishes, particularly affecting light deflection and the stability of photon orbits. Through both weak and strong deflection analyses, this study indicates the need for ultrasensitive technology to detect these higher-dimensional signatures. Remarks on the observational data constraints currently favor four-dimensional spacetime; however, the exploration of additional dimensions remains vital in advancing models of quantum gravity. This work provides a theoretical framework for understanding black hole behavior in higher dimensions, potentially informing future astrophysical observations.
%We investigate the impact of higher dimensions on the properties of Schwarzschild-Tangherlini black holes, focusing on the photonsphere, black hole shadow, deflection angles, and quasinormal modes (QNMs). We find that these properties diminish as the dimensionality ($n$) of the black hole increases. Analysis of the shadow radius measured by the Event Horizon Telescope suggests non-integer dimensions around $n\lessgtr4$. We derive an analytic formula for the weak field deflection angle, highlighting the need for advanced sensitive detection devices to observe lensed images influenced by higher dimensions. Our study of QNMs using physics-informed neural networks and the WKB method reveals a convergence towards known relationships between QNM frequencies and photon-sphere orbit frequencies. Despite the energetic nature of perturbing fields in higher dimensions, their damping increases. This suggests a complex interplay between dimensionality and the dynamics of black hole phenomena.
\end{abstract}

%only 5 keywords 

\maketitle

%\tableofcontents

%%%%%%%
\section{Introduction}\label{intr}
Black holes, enigmatic cosmic entities born from the gravitational collapse of massive stars, have long captivated the imagination of scientists and the general public alike. These celestial phenomena serve as unique laboratories, pushing the boundaries of our understanding of gravity and the very fabric of spacetime. Black holes become crucial probes for testing the predictions of Einstein's general theory of relativity, unraveling the mysteries of galactic evolution, and offering insights into the larger cosmic landscape \cite{Penrose:1964wq,Hawking:1974rv}. In this pursuit, black holes are indispensable tools for verifying the robustness of general relativity in extreme conditions. The Event Horizon Telescope's (EHT) imaging of the supermassive black hole in the M87 galaxy in 2019 provides a compelling visual confirmation of theoretical predictions concerning event horizons and photonspheres \cite{EventHorizonTelescope:2019dse,EventHorizonTelescope:2019ths, EventHorizonTelescope:2022xqj}, which affirms the seminal works of Synge, Luminet, and Falcke \cite{Synge:1966okc, Luminet:1979nyg,Falcke:1999pj}. Another success immediately followed in 2022, as the black hole at the heart of our galaxy was confirmed \cite{EventHorizonTelescope:2022wkp,EventHorizonTelescope:2022wok}. In addition to the shadow detection, gravitational waves were detected by LIGO and Virgo from black hole mergers has offered unprecedented evidence supporting general relativity's predictions about spacetime dynamics \cite{LIGOScientific:2016aoc,LIGOScientific:2017vwq,KAGRA:2022osp}, and opening a new era in astronomy.

As such, in exploring black holes beyond the confines of our familiar four-dimensional spacetime, we are compelled by several reasons: (1) Theoretical frameworks like string theory propose the existence of extra spatial dimensions beyond the familiar three \cite{Randall:1999vf,Maldacena:1997re}. Black holes in higher dimensions emerge naturally from these theories, and are often invoked in discussions of quantum gravity (See Ref. \cite{Emparan:2008eg} for a comprehensive review). Studying Schwarzschild black holes in extra-dimensions helps probe the quantum nature of spacetime and gravity, potentially providing clues about the microscopic structure of black holes; (2) Higher-dimensional gravity theories, such as Kaluza-Klein theory, aim to unify gravity with other fundamental forces. Black holes in higher dimensions are key in testing and understanding these unified theories; (3) Some theoretical models involve the concept of higher-dimensional spacetimes with branes representing our visible universe. Black holes in these scenarios can have unique properties and effects that differ from their counterparts in standard four-dimensional spacetime; and (4) Studying black holes in higher dimensions is not only motivated by theoretical physics but also by mathematical interest. Higher-dimensional spacetimes can reveal new insights and challenges in understanding geometry and general relativity. These motivations, led to several studies present in the literature \cite{Emparan:2008eg,Myers:1986un,Cadeau:2000tj,Shen:1989uj,Iyer:1989nd,Chodos:1980df,Xu:1988ju,Cai:2020igv,Do:2019jpg,Feng:2015jlj,Myung:2013cna,Paul:2023gzj} as the scientific community tries to uncover the effects of higher dimensions on certain aspects of the black hole geometry. Investigating black holes in extra dimensions may help identify unique signatures that could be observed or tested through astrophysical observations such as in the black hole shadow \cite{Papnoi:2014aaa, Singh:2017vfr}, gravitational lensing and optical properties \cite{Abdujabbarov:2015rqa,Belhaj:2020rdb}, quasinormal modes \cite{Matyjasek:2021xfg, Yan:2020hga}, thermodynamics \cite{Andre:2021ctu, Chen:2008ra,Barman:2019vst}, etc \cite{Fox:2019zzn, Kunstatter:2013aqa, Ishibashi:2003ap, Ahmedov:2021ohg,Lake:2003gi}.

It is understood that the higher dimensionality of black holes is theoretically and mathematically appealing. With the recent constraints for the radius of the invisible shadow provided by the EHT, our aim in this paper is to constrain the dimensionality of the black hole and whether some of the uncertainties in the shadow radius may be caused by the higher dimensionality of the black hole. Furthermore, we explore the deflection angles within the weak and strong field regimes, checking higher dimensional signatures. To study the weak field deflection angle, we utilize the Gauss-Bonnet theorem to include the finite distance and time-like particles \cite{Gibbons:2008rj,Ishihara:2016vdc,Li:2019vhp,Li:2020dln,Li:2020wvn}, to give us the general case.

The strong deflection angle, commonly denoted as the angle of light bending in the proximity of a massive object, encapsulates the profound gravitational influence on the trajectory of light rays. Unlike weak deflection, characterized by minimal light deviation, strong deflection occurs when light traverses in close proximity to the gravitational source, resulting in significant bending of light rays. This phenomenon manifests observable signatures that offer invaluable insights into the nature and characteristics of the intervening mass \cite{Virbhadra:1999nm,Virbhadra:2002ju,Virbhadra:2008ws, Virbhadra:2022iiy, Virbhadra:2007kw}. Note that the works of Bozza \cite{Bozza:2002zj} and Tsukamoto \cite{Tsukamoto:2016jzh} have investigated the calculation of the strong deflection angle in the vicinity of a black hole. Tsukamoto, in particular, extends Bozza's methodology by introducing a novel term, denoted as $z$, enabling an analytical approximation for the deflection angle equation. The outcomes exhibit consistency, with negligible discrepancies \cite{Tsukamoto:2016jzh}. Tsukamoto's approach, and its subsequent refinement, broaden the applicability of the strong deflection angle formalism to encompass various black hole models, thus facilitating investigations into more intricate black hole configurations \cite{Tsukamoto:2021caq, Tsukamoto:2020iez, Tsukamoto:2021apr}. Furthermore, numerous insightful findings have been presented concerning the behavior of light under such conditions \cite{Wang:2019cuf, Chagoya:2020bqz, Tsukamoto:2021caq, Tsukamoto:2022uoz, Abbas:2021whh, Belhaj:2020rdb,Virbhadra:2024xpk,Hsieh:2021rru,Hsieh:2021scb}. For instance, within the strong field regime, the influence of charged matter and dark matter on the deflection angle becomes more discernible in comparison to weak deflection scenarios \cite{Lobos:2022jsz}.

The formulation of theories elucidating the strong deflection angle within higher-dimensional black hole spacetimes holds profound implications for our understanding of black hole physics \cite{Tsukamoto:2014dta}. It not only sheds light on the behavior of gravity within unconventional higher-dimensional contexts but also elucidates fundamental principles underlying black hole theory \cite{Friedlander:2023qmc}. The exploration of how gravitational lensing attributes are influenced by higher dimensions possesses the potential to refine our conceptual frameworks regarding black hole formation, evolution, and the interpretation of observational data pertaining to astrophysical phenomena involving black holes. As we shall discuss, the photonsphere plays an important role in forming the black hole shadows, as it dictates the size and shape of the black hole shadow. However, complementary information can be extracted by studying quasinormal modes (QNMs) of black holes. QNMs are characteristic oscillations of black holes that arise when the black hole is perturbed. These perturbations can be thought of as the ``ringing'' of the black hole, and they decay over time due to the emission of gravitational waves. QNMs are defined by complex frequencies, where the real part represents the oscillation frequency and the imaginary part denotes the damping rate. The link between photon orbits and QNMs is established through the concept that the real part of the quasinormal mode frequencies is related to the orbital frequency of the photonsphere. In contrast, the imaginary part is associated with the Lyapunov exponent, which characterizes the stability of the photon orbits. This relationship was demonstrated by Cardoso et al. \cite{Cardoso:2008bp}, who showed that for high-frequency (eikonal) QNMs, the frequencies are directly tied to the properties of the photonsphere. The oscillation frequencies of QNMs therefore correspond to the orbital frequencies of photons in the photonsphere, and the decay rates correspond to the stability properties of these orbits. As noted in Refs. \cite{KONOPLYA2023137674, KONOPLYA2017597} some care is required when calculating QNMs in extra-dimensional spacetimes and presupposing this correspondence in the eikonal regime, which has been demonstrated to be invalid whenever the BH perturbation equations have effective potentials that are not ``well-behaved'' from the stand-point of applicability of the WKB method. Inferring from the effective potentials of asymptotically flat Schwarzschild-Tangherlini BHs, which have a single extremum in the domain $r \in [2M, \infty)$, the WKB method (and, in turn, the correspondence) is guaranteed to be applicable in this case. But in order to move toward a full analysis of the phenomena, we consider here the behavior of the deflection angle in the presence of a higher-dimensional black hole, and how to calculate QNMs in such a space in a new robust neural network manner.

We organize the paper as follows: In Sect. \ref{sec2}, we provide a quick review of the Schwarzschild-Tanghelini black hole, along with examining the higher dimension parameter $n$ to the photon orbit, and the black hole shadow. Next, we study the weak and strong deflection angles of this higher-dimensional black hole in Sect. \ref{sec3}. Finally, we employed machine learning methods to calculate the QNMs in Sect. \ref{sec4} using PINN approximations. We give concluding remarks and future research prospects in Sect. \ref{conc}. Throughout this paper, we used $(-,+,+,+)$ metric signature and geometrized units by considering $G = c = 1$.

%%%%%%%
\section{Shadow of a Schwarzschild black hole in higher dimensions}\label{sec2}
The Schwarzschild-Tangherlini metric extends the Schwarzschild solution to higher dimensions in the context of general relativity. It provides a solution to the Einstein field equations for a spherically symmetric, non-rotating black hole in a spacetime with $n$ dimensions, where $n$ is greater than $4$. For its full review, see Ref. \cite{Emparan:2008eg}, where the metric of the Schwarzschild black hole in extra-dimensions (Schwarzschild-Tangherlini black hole) is given by
\begin{equation} \label{n_metric}
    ds^2 = -A(r) dt^2 + B(r) dr^2 + r^2 d\Omega^2_{n-2},
\end{equation}
where $A(r)$, $B(r)$, and $d\Omega^2_{n-2}$ are expressed by
\begin{align}\label{mfunction}
    A(r) &= 1- \frac{2M}{r^{n-3}}, \nonumber \\
    B(r) &= A(r)^{-1}, \nonumber \\
    d\Omega^2_{n-2} &= d\theta_1^2 + \sin^2(\theta_1) d\theta_2^2 + ... + \prod\limits_{i=1}^{n-3} \sin^2(\theta_i) d\theta_{n-2}^2 ,
\end{align}
respectively. As one considers only the pure Schwarzschild field and its spherical symmetry, it can be noted from Ref. \cite{Tangherlini:1963bw} that $d\theta_1 = d\theta = 0$, $d\theta_2 = d\phi$, and $d\theta_{n-2} = 0$. Furthermore, $M$ admits the dimensionality $n$ through
\begin{equation}
    M = \frac{8\pi m}{(n-2)\Omega_{n-2}},
\end{equation}
where $\Omega_{n-2}$ is the area of the $(n-2)-$dimensional unit sphere: 
\begin{equation}
    \Omega_{n-2} = \frac{2\pi^{\left(\frac{n-1}{2} \right)}}{\Gamma{\left(\frac{n-1}{2} \right)}},
\end{equation}
and $m$ is the black hole mass.

The photonsphere $(r_{\rm ps})$ is a region around a black hole where photons can theoretically orbit the black hole in a circular path. It is well known that for the Schwarzschild case, $r_{\rm ps} = 3m$. This section shows how the number of dimensions $n$ affects the photonsphere, as we apply the formalism developed in Ref. \cite{Perlick:2021aok}. The results here will be important in the next section.

\begin{figure*}
    \centering
    \includegraphics[width=0.48\textwidth]{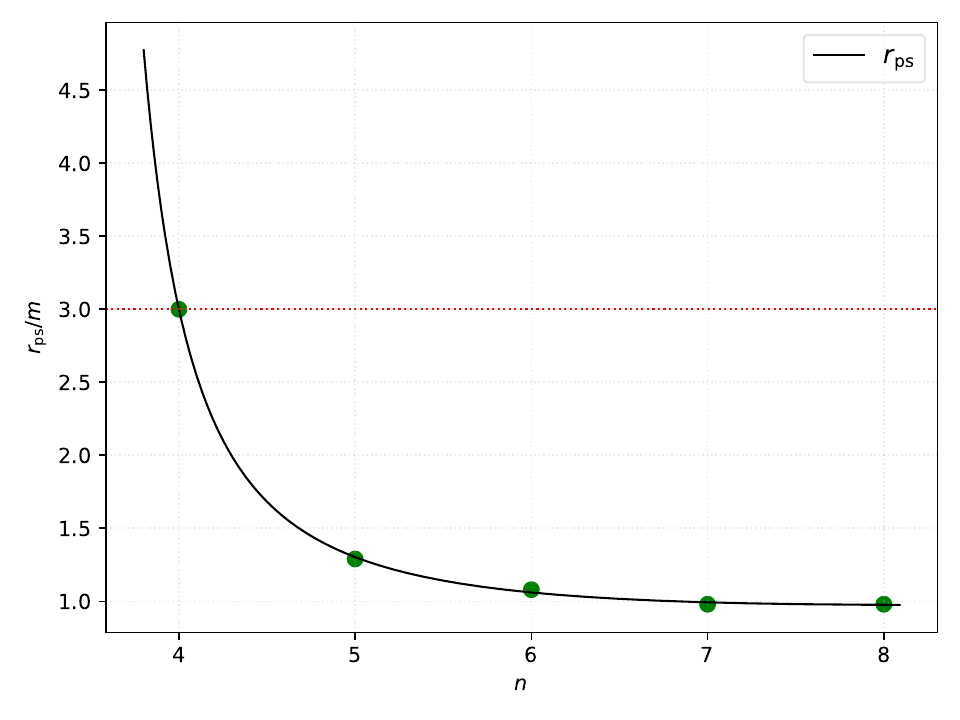}
    \includegraphics[width=0.48\textwidth]{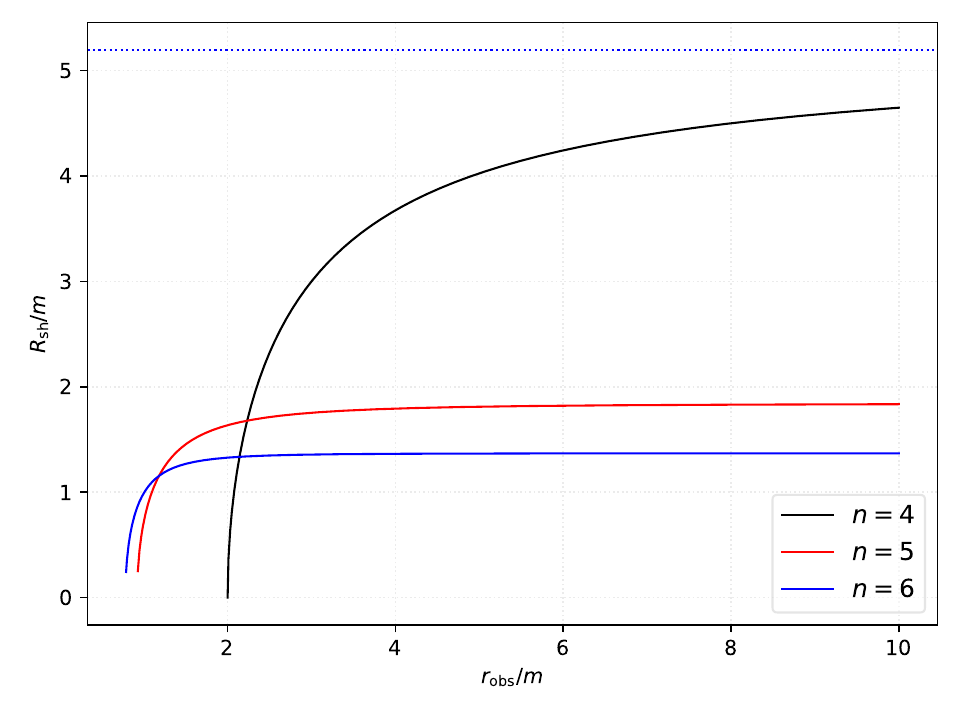}
    \caption{The left panel shows the influence of $n$ on the photonsphere radius. The right panel shows the influence of $n$ and $r_{\rm obs}$ on the shadow radius. The horizontal dotted line is the classical Schwarzschild case.}
    \label{rps}
\end{figure*}

We take the standard Lagrangian
\begin{equation}
    \mathcal{L} = \frac{1}{2}\left( -A(r) \dot{t}^2 +B(r) \dot{r}^2 + r^2 \dot{\phi}^2 \right).
\end{equation}
Applying the variational principle gives us the two constants of motion:
\begin{equation}
    E = A(r)\frac{dt}{d\lambda}, \qquad L = r^2\frac{d\phi}{d\lambda},
\end{equation}
where the impact parameter can be defined as
\begin{equation}
    b \equiv \frac{L}{E} = \frac{r^2}{A(r)}\frac{d\phi}{dt}.
\end{equation}
Light rays obey $g_{\mu \nu}\dot{x}^\mu \dot{x}^\nu = 0$, leading to an orbit equation
\begin{equation}
\label{eq.8}
    \left(\frac{dr}{d\phi}\right)^2 =\frac{r^2}{B(r)}\left(\frac{h(r)^2}{b^2}-1\right),
\end{equation}
where the function $h(r)$ is defined as \cite{Perlick:2021aok}
\begin{equation}
     h(r)^2 = \frac{r^2}{A(r)}.
\end{equation}
The location of the photonsphere can be found by either satisfying the condition $dr/d\phi = d^2r/d\phi^2 = 0,$ or using $h'(r) = 0$, which can lead to
\begin{equation}
    A'(r)r^2 - 2rA(r) = 0,
\end{equation}
resulting to
\begin{equation} \label{eq_rps}
    2r - \frac{16 m \pi^{\frac{3-n}{2}} \Gamma \! \left(\frac{n+1}{2}\right) r^{4-n} }{n - 2}=0.
\end{equation}
The equation above makes it possible to normalize $m$ so that $r$ becomes dimensionless, which makes it convenient for plotting purposes. Solving the above for $r$ yields the photonsphere radius
\begin{equation} \label{ana_rps}
    r_{\rm ps} = \exp{\left[\frac{2 \ln \! \left(\Gamma \! \left(\frac{n+1}{2}\right)\right)-2 \ln \! \left(n -2\right)+\left(3-n\right) \ln \! \left(\pi \right)+6 \ln \! \left(2\right)}{2 (n -3)}\right]}.
\end{equation}
The numerically plot of $r_{\rm ps}$ is shown in the left panel of Fig. \ref{rps}, where we can see the immediate effect of increasing $n$ to the photonsphere radius $r_{\rm ps}$.

%%%%%%%
The presence of the photonsphere plays a crucial role in shaping the appearance of the black hole silhouette and contributes to the distinctive features observed in images, such as the shadow or the bright ring surrounding it. After obtaining the location of the photonsphere through the analytic equation \eqref{ana_rps}, a small perturbation can either make the photons escape or spiral toward the black hole. Escaped photons will contribute to the black hole silhouette observed at $r = r_{\rm obs}$, which can backward trace the photons' path to study the black hole shadow formation. For such an observer \cite{Perlick:2021aok},
\begin{equation}
    \tan(\alpha_{\text{sh}}) = \lim_{\Delta x \to 0}\frac{\Delta y}{\Delta x} = \left(\frac{r^2}{B(r)}\right)^{1/2} \frac{d\phi}{dr} \bigg|_{r=r_\text{obs}},
\end{equation}
which can be simplified as
\begin{equation}
    \sin^{2}(\alpha_\text{sh}) = \frac{b_\text{crit}^{2}}{h(r_\text{obs})^{2}},
\end{equation}
where the critical parameter is defined by
\begin{equation}
    b_{\rm crit}^2 = \frac{r_{\rm ps}^2}{A(r_{\rm ps})}.
\end{equation}
With the help of Eq. \eqref{ana_rps}, the exact analytic expression for the shadow radius is then
\begin{align} \label{ana_rsh}
    R_\text{sh} &= b_\text{crit}\sqrt{A(r_\text{obs})} \\
    &=\Bigg\{\frac{n-1}{n-3}\exp{\left[\frac{2 \ln \! \left(\Gamma \! \left(\frac{n+1}{2}\right)\right)-2 \ln \! \left(n -2\right)+\left(3-n\right) \ln \! \left(\pi \right)+6 \ln \! \left(2\right)}{n -3}\right]}\left(1-\frac{8 \pi^{\frac{3-n}{2}} \Gamma \! \left(\frac{n-1}{2}\right) r_{\rm obs}^{3-n}}{n -2} \right)\Bigg\}^{1/2}.
\end{align}
One can check that black hole shadow formation for $n=1$ up to $n=3$ is not allowed. In the right panel of Fig. \ref{rps}, we plot Eq. \eqref{ana_rsh}, where we can see the effect of both $n$ and $r_{\rm obs}$. The black curve describes the standard $n = 4$ Schwarzschild case. We see how the rate at which $R_\text{sh}$ changes with $r_\text{obs}$ is high near the black hole, and how it becomes zero as $r_\text{obs} \to \infty$. Far smaller shadows can be observed for $n = 5$ and $n = 6$, and such a distinction from $n = 4$ can be seen when $r_\text{obs} \to \infty$. Furthermore, there are intersection points of the curves due to different $n$. For instance, the distinction between $n = 4$ and $n = 5$ is gone when the observer is near the black hole. However, from a physical standpoint, the shadow radius will be smaller than the event horizon, which is not possible. As a final remark, while Eq. \eqref{ana_rsh} is quite a complicated expression, we can approximate the equation as $r_{\rm obs} \to \infty$. The following agrees with the plot we obtained:
\begin{align}
    R_{\rm sh} = 
\begin{cases} 
3 \sqrt{3}-\frac{3 \sqrt{3}}{r_{\rm obs}} & \text{if } n = 4, \\ 
\frac{4 \sqrt{6}}{3 \sqrt{\pi}}-\frac{16 \sqrt{6}}{9 \pi^{3/2} r_{\rm obs}^{2}} & \text{if } n = 5, \\
\frac{2^{1/3} 15^{5/6}}{6 \pi^{1/3}}-\frac{2^{1/3} 15^{5/6}}{8 \pi^{4/3} r_{\rm obs}^{3}} & \text{if } n = 6.
\end{cases}
\end{align}

\begin{comment}
\begin{table}
    \centering
    \begin{tabular}{ p{2cm} p{3.5cm} p{4.5cm} p{2cm}}
    \hline
    \hline
    Black hole & Mass $m$ ($M_\odot$) & Angular diameter: $2\alpha_\text{sh}$ ($\mu$as) & Distance (kpc) \\
    \hline
    Sgr. A*   & $4.3 \pm 0.013$x$10^6$ (VLTI)    & $48.7 \pm 7$ (EHT) &   $8.277 \pm 0.033$ \\
    M87* &   $6.5 \pm 0.90$x$10^9$  & $42 \pm 3$   & $16800$ \\
    \hline
    \end{tabular}
    \caption{Black hole observational constraints as derived from \cite{EventHorizonTelescope:2019dse}.}
    \label{tab1}
\end{table}

\begin{figure*}
    \centering
    \includegraphics[width=0.48\textwidth]{sha.pdf}
    \caption{Left: Dependence of shadow radius to the observer's location for different values of $n$. Right: Shadow constraints in $n$ with a simple parameter estimation using the EHT data. The blue horizontal dotted line is the Schwarzschild shadow $R_{\rm sh}/M = 3\sqrt{3}$.}
    \label{sha}
\end{figure*}

\end{comment}

%%%%%%%
\section{Particle deflection in weak and strong field regime}\label{sec3}

\subsection{Weak deflection angle}
In this section, we aim to analyze the effect of the number of dimensions on the weak deflection angle. To do so, we use the Gauss-Bonnet theorem stating that \cite{Carmo2016,Klingenberg2013}
\begin{equation} \label{eGBT}
    \iint_D KdS+\sum\limits_{i=1}^N \int_{\partial D_{a}} \kappa_{\text{g}} d\ell+ \sum\limits_{i=1}^N \Theta_{i} = 2\pi\chi(D).
\end{equation}
Here, $K$ is the Gaussian curvature, $dS$ is the area measure, $\Theta_i$, and $\kappa_g$ is the jump angles and geodesic curvature of $\partial D$, respectively, and $d\ell$ is the arc length measure. The application to null geodesics in the equatorial plane implies that the Euler characteristics should be $\chi(D) = 1$. If the integral is evaluated over the infinite area surface bounded by the light ray, it was shown in Ref. \cite{Ishihara:2016vdc} that the above reduces to
\begin{equation} \label{eIshi}
    \hat{\alpha}=\phi_{\text{RS}}+\Psi_{\text{R}}-\Psi_{\text{S}} = -\iint_{_{\text{R}}^{\infty }\square _{\text{S}}^{\infty}}KdS,
\end{equation}
where $\hat{\alpha}$ is the weak deflection angle. In the above formula, $\phi_\text{RS} = \phi_\text{R} - \phi_\text{S}$ is the azimuthal separation angle between the source S and receiver R, $\phi_\text{R}$ and $\phi_\text{S}$ are the positional angles, and $_\text{R}^{\infty }\square _\text{S}^{\infty}$ is the integration domain. While this study can use the above formula, we prefer to use its extension to the non-asymptotically flat case given in Ref. \cite{Li:2020wvn}, where one uses the path in the photonsphere orbit instead of the path at infinity. Eq. \eqref{eIshi} can be reformulated as:
\begin{equation} \label{eLi}
    \hat{\alpha} = \iint_{_{r_\text{ps}}^{R }\square _{r_\text{ps}}^{S}}KdS + \phi_{\text{RS}}.
\end{equation}

Before using the above equation, we are interested in the deflection angle of massive particles. Such a general case reduces to a special case of photon deflection when $v = 1$. As such, we need the Jacobi metric stating that 
\begin{align} \label{eJac}
    dl^2=g_{ij}dx^{i}dx^{j}
    =(E^2-\mu^2A(r))\left(\frac{B(r)}{A(r)}dr^2+\frac{r^2}{A(r)}d\phi^2\right),
\end{align}
where $E$ is the energy per unit mass ($\mu$) of the massive particle:
\begin{equation} \label{en}
    E = \frac{\mu}{\sqrt{1-v^2}}.
\end{equation}
It is then useful to define another constant quantity in terms of the impact parameter $b$, which is the angular momentum per unit mass:
\begin{equation}
    J = \frac{\mu v b}{\sqrt{1-v^2}},
\end{equation}
and with $E$ and $J$, we can define the impact parameter as
\begin{equation}
	b = \frac{J}{vE}.
\end{equation}
Using the line element for time-like particles $ds^2=g_{\mu \nu}dx^{\mu}dx^{\nu} = -1$ and defining $u = 1/r$, the orbit equation can be derived as
\begin{align} \label{eorb}
    F(u) \equiv \left(\frac{du}{d\phi}\right)^2 
    = \frac{1}{A(u)B(u)}\Bigg[\left(\frac{1}{vb}\right)^2-A(u)\left(\frac{1}{J^2}+u^2\right)\Bigg],
\end{align}
which, in our case, yields ($p=n-3$)
\begin{equation} \label{eorb2}
    F(u)=\frac{1}{v^{2}b^2}+\left(\frac{1}{J^{2}}+u^{2}\right)\left[\frac{2M\,u^p \left(u^{2} J^{2}+1\right)}{J^{2}}\right].
\end{equation}
Next, by an iterative method, the goal is to find $u$ as a function of $\phi$, which we find as
\begin{equation} \label{euphi}
    u(\phi) = \frac{\sin(\phi)}{b}+\frac{1+v^2\cos^2(\phi)}{b^2v^2}M.
\end{equation}
From here, we obtain the solution for $\phi$ as
\begin{equation} \label{ephi}
    \phi =\arcsin(bu) + \frac{M\left[v^{2}\left(b^{2}u^{2}-1\right)-1\right]}{bv^{2}\sqrt{1-b^{2}u^{2}}}.
\end{equation}
This is the expression for the positional angle $\phi_\text{S}$, which implies that $u \rightarrow u_\text{S}$. Note also that $\phi_R = \pi -\phi_S$.

The Gaussian curvature can be derived using
\begin{align}
    K=-\frac{1}{\sqrt{g}}\left[\frac{\partial}{\partial r}\left(\frac{\sqrt{g}}{g_{rr}}\Gamma_{r\phi}^{\phi}\right)\right],
\end{align}
since $\Gamma_{rr}^{\phi} = 0$ from Eq. \eqref{eJac}. Furthermore, the determinant of Eq. \eqref{eJac} is
\begin{equation}
    g=\frac{B(r)r^2}{A(r)^2}(E^2-\mu^2 A(r))^2.
\end{equation}
With the analytical solution to $r_\text{ps}$, it is easy to see that
\begin{equation}
    \left[\int K\sqrt{g}dr\right]\bigg|_{r=r_\text{ps}} = 0,
\end{equation}
which yields
\begin{align} \label{gct}
    \int_{r_\text{ps}}^{r(\phi)} K\sqrt{g}dr &= -\frac{A(r)\left(E^{2}-A(r)\right)(2r)-E^{2}r^2A(r)'}{2r A(r)\left(E^{2}-A(r)\right)\sqrt{B(r)}}\bigg|_{r = r(\phi)} \nonumber \\
    &=-1 + \frac{M \left[-1+\left(p +1\right) E^{2}\right]}{E^{2}-1} \left[ b^{-p} \left(\frac{1}{\sin \! \left(\phi \right)}\right)^{-p} \right] ,
\end{align}
where the prime denotes differentiation with respect to $r$. The weak deflection angle is then \cite{Li:2020wvn},
\begin{align} \label{eqwda}
    \hat{\alpha} = \int^{\phi_\text{R}}_{\phi_\text{S}} \left[-\frac{A(r)\left(E^{2}-A(r)\right)(2r)-E^{2}r^2A(r)'}{2rA(r)\left(E^{2}-A(r)\right)\sqrt{B(r)}}\bigg|_{r = r(\phi)}\right] d\phi + \phi_\text{RS}.
\end{align}
Using Eq. \eqref{euphi} in Eq. \eqref{gct}, we find
\begin{align} \label{gct2}
    &\left[\int K\sqrt{g}dr\right]\bigg|_{r=r_\phi} =-\phi_\text{RS} -\left(\frac{M \left[\left(p +1\right) E^{2} - 1\right]}{E^{2}-1} \right) b^{-p} \cos \! \left(\phi \right) \left(\sin^{2}\left(\phi \right)\right)^{\frac{1}{2}-\frac{p}{2}} \left(\csc^{1-p}\left(\phi \right)\right) {}_2F_{1} \left(\frac{1}{2},\frac{1}{2}-\frac{p}{2};\frac{3}{2};\cos^{2}\left(\phi \right)\right)\bigg|_{\phi_S}^{\phi_R} \nonumber \\
    &=-\phi_\text{RS} + \left(\frac{2 M \left[\left(p +1\right) E^{2} - 1\right]}{E^{2}-1} \right) b^{-p} \cos \! \left(\phi \right) \left(\sin^{2}\left(\phi \right)\right)^{\frac{1}{2}-\frac{p}{2}} \left(\csc^{1-p}\left(\phi \right)\right) {}_2F_{1}\left(\frac{1}{2},\frac{1}{2}-\frac{p}{2};\frac{3}{2};\cos^{2}\left(\phi \right)\right)
\end{align}
since $\phi_\text{S}$ has the same form as Eq. \eqref{ephi}, while keeping in mind that the definite integral is already evaluated. Then, we find Eq. \eqref{eqwda} as
\begin{equation} \label{ewda2}
    \hat{\alpha} = \left(\frac{2 M \left[\left(p +1\right) E^{2} - 1\right]}{E^{2}-1} \right) b^{-p} \cos \! \left(\phi \right) \left(\sin^{2}\left(\phi \right)\right)^{\frac{1}{2}-\frac{p}{2}} \left(\csc^{1-p}\left(\phi \right)\right) {}_2F_{1}\left(\frac{1}{2},\frac{1}{2}-\frac{p}{2};\frac{3}{2};\cos^{2}\left(\phi \right)\right).
\end{equation}
Finally, we need to express the equation above in terms of finite distance $u_\text{S}$ and $u_\text{R}$. We use the following relations:
\begin{align} \label{ecos}
    &\cos(\phi) = \sqrt{1-b^{2}u^{2}}-\frac{M u\left[v^{2}\left(b^{2}u^{2}-1\right)-1\right]}{v^2\sqrt{\left(1-b^{2}u^{2}\right)}}, \nonumber\\
    &\left(\sin^{2}\left(\phi \right)\right)^{\frac{1}{2}-\frac{p}{2}} = \frac{b u}{(bu)^{p}} - \frac{M \left(b^{2} u^{2}\right)^{\frac{1}{2}-\frac{p}{2}} \left[v^{2} \left(b^{2} u^{2}-1\right)- 1\right] \left(p-1 \right)}{u \,v^{2} b^{2}}, \nonumber \\
    & \csc^{1-p}\left(\phi \right) = \frac{(bu)^{p}}{b u} + \frac{\left[v^{2} \left(b^{2} u^{2}-1\right)- 1\right] \left(p-1 \right) \left(\frac{1}{b u}\right)^{-p} M}{b^{3} u^{2} v^{2}}.
\end{align}
After approximating Eq. \eqref{ewda2} again, the general analytic expression for the deflection angle of the Schwarzschild-Tangherlini black hole with finite distance as
\begin{equation} \label{gen_wda}
    \hat{\alpha} = \left(1+\frac{p}{v^{2}}\right) M \,b^{-p}\left[\sqrt{1-b^{2} u_\text{S}^{2}}\, {}_2F_{1} \left(\frac{1}{2},\frac{1}{2}-\frac{p}{2};\frac{3}{2};-b^{2} u_\text{S}^{2}+1\right) + \sqrt{-b^{2} u_\text{R}^{2}+1}\, {}_2F_{1} \left(\frac{1}{2},\frac{1}{2}-\frac{p}{2};\frac{3}{2};-b^{2} u_\text{R}^{2}+1\right) \right],
\end{equation}
which is written in terms of a hypergeometric function.

\begin{figure*}
    \centering
    \includegraphics[width=0.48\textwidth]{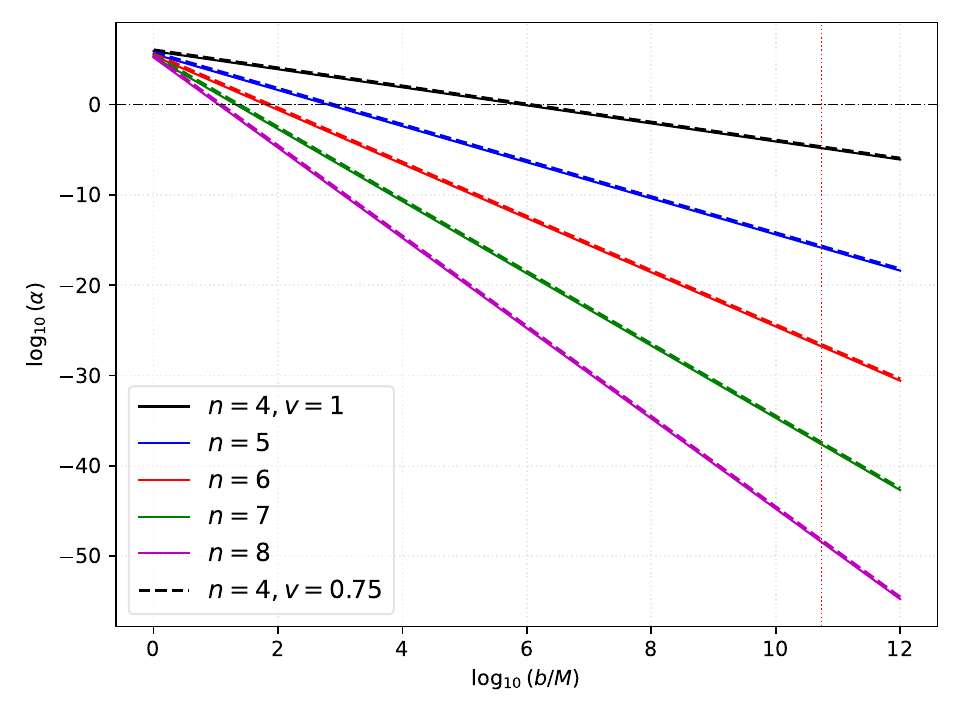}
    \includegraphics[width=0.48\textwidth]{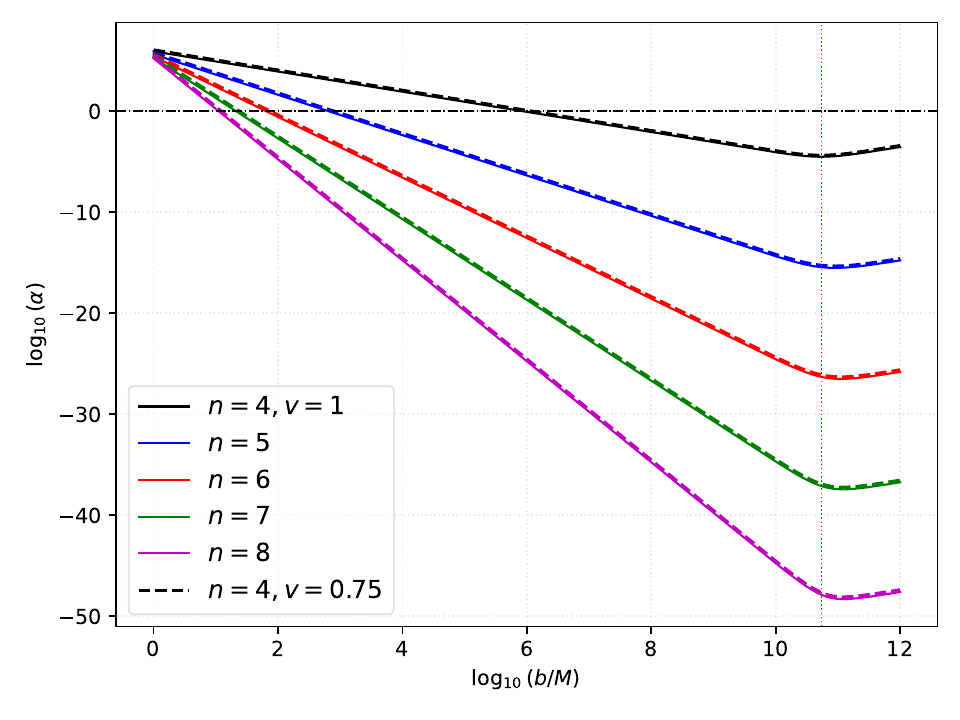}
    \caption{The behavior of $\alpha$ (in $\mu$as) for various dimensions $n$. Left panel: $u/M=0$. Right panel: $u/M = 1.85\times 10^{-11}$, corresponding to the reciprocal of our distance from M87* (also represented by the red vertical dotted line). The other dashed line represents the time-like case from $n=5$ to $n=8$.}
    \label{wda}
\end{figure*}

One can easily check that if $p=1$ (where $n=4$), the equation above reduces to the timelike expression in finite distance as
\begin{equation}
    \hat{\alpha}^{n=4}_{\rm timelike} = \frac{\left(v^{2}+1\right) m}{b \,v^{2}} \left[ \sqrt{1-b^{2} u_\text{S}^{2}} + \sqrt{1-b^{2} u_\text{R}^{2}} \right].
\end{equation}
If we assume that $u_\text{S} = u_\text{R}$ and both are approximately zero, then
\begin{equation}
    \hat{\alpha}^{n=4}_{\rm timelike} = \frac{2\left(v^{2}+1\right) m}{b \,v^{2}}.
\end{equation}
Finally, for photons where $v = 1$,
\begin{equation}
    \hat{\alpha}^{n=4}_{\rm photon} = \frac{4m}{b}.
\end{equation}
We observe that when $n$ is odd, $\hat{\alpha}$ contains a factor involving a hypergeometric function. For instance, when $n = 5$ ($p=2$),
\begin{equation}
    \hat{\alpha}^{n=5}_{\rm timelike} = \frac{\left(v^{2}+2\right) M}{b^{2} v^{2}} \left[\sqrt{1-b^{2} u_\text{S}^{2}}\, {}_2F_{1} \left(-\frac{1}{2},\frac{1}{2};\frac{3}{2};-b^{2} u_\text{S}^{2}+1\right) + \sqrt{-b^{2} u_\text{R}^{2}+1}\, {}_2F_{1} \left(-\frac{1}{2},\frac{1}{2};\frac{3}{2};-b^{2} u_\text{R}^{2}+1\right) \right].
\end{equation}
When $n$ is even, such as when $p = 3$, we observe that $\hat{\alpha}$ does not contain a factor involving a hypergeometric function:
\begin{equation}
    \hat{\alpha}^{n=6}_{\rm timelike} = \frac{\left(v^{2}+3\right) M}{3 b^{3} v^{2}} \left[ \sqrt{1-b^{2} u_\text{S}^{2}}\, \left(2+b^{2} u_\text{S}^{2}\right) + \sqrt{1-b^{2} u_\text{R}^{2}}\, \left(2+b^{2} u_\text{R}^{2}\right)\right].
\end{equation}
We plot Eq. \eqref{gen_wda}, shown in Fig. \ref{wda}. First, we observe that the deflection angle caused by time-like particles gives a slightly higher value of $\alpha$ than photon deflection. Second, our observation is that as the number of dimensions increases, $\alpha$ decreases. It only means that if we use the phenomenon of deflection angle to probe the existence of higher dimensions in black holes, more sensitive devices are needed. Finally, the effect of finite distance indicates that $\alpha$ slightly increases as the impact parameter $b/M$ becomes greater than the observer distance from the black hole.

\subsection{Strong Deflection Angle}
Photons slightly greater than the photonsphere radius will undergo the phenomenon of strong deflection angle. In this section, we calculate and showcase the deflection of light as it approaches the photonsphere in the strong field limit. Using the approach of Tsukamoto in Ref. \cite{Tsukamoto:2016jzh}, the deflection angle is derived  using the light trajectory shown in Eq. \eqref{eq.8} but, in this section we express it as,
\begin{equation}
\label{eq.43}
\left(\frac{dr}{d\phi}\right)^{2} = \frac{R(r)r^2}{B(r)},
\end{equation}
where
\begin{equation}
\label{eq.44}
R(r) = \frac{A({r_0})r^2}{A(r)r_{0}^2}-1.
\end{equation}
Note that the $A(r)$ is the metric function defined by Eq. \eqref{mfunction}, while $A(r_{0})$ is the metric function evaluated at distance $r_{0}$. The solution of Eq. \eqref{eq.43} yields the strong deflection angle $\alpha(r_{0})$ as shown in Ref. \cite{Tsukamoto:2016jzh, Bozza:2002zj, Virbhadra:1998dy},  
\begin{equation}
\begin{split}
\label{eq.45}
\alpha(r_{0}) &= I(r_{0}) - \pi \\
&= 2 \int^{\infty}_{r_{0}} \frac{dr}{\sqrt{\frac{R(r)C(r)}{B(r)}}}-\pi.
\end{split}
\end{equation}
In order to evaluate the integral in Eq. \eqref{eq.45}, we expand over $r=r_{0}$. This yields a regular integral $\kappa_{R}$ and a diverging integral $\kappa_{D}$, and by introducing a new variable, z, defined as, 
\begin{equation}
z \equiv 1-\frac{r_{0}}{r},
\end{equation}
$I(r_{0})$ is expressed as,
\begin{equation}
I(r_{0}) = \int^{1}_{0}\kappa(z, r_{0})dz = \int^{1}_{0}\left[\kappa_{D}(z, r_{0})+\kappa_{R}(z, r_{0})\right]dz,
\end{equation}
where $\kappa(z, r_{0})$ is expressed as the sum of the diverging integral, $\kappa_{D}$, and regular integral, $\kappa_{R}$. The details of the expansion of Eq. \eqref{eq.45} was shown in Refs. \cite{Tsukamoto:2016jzh,Bozza:2002zj}. As a result the strong deflection angle is expressed as, 
\begin{equation}
\label{eq.48}
\hat{\alpha}_{\text{str}} = -\bar{a} \log \left(\frac{b_0}{b_\text{crit}}-1\right)+\bar{b}+O\left(\frac{b_{0}}{b_{c}}-1\right)\log\left(\frac{b_{0}}{b_{c}}-1\right),
\end{equation}
where $\bar{a}$ and $\bar{b}$ are coefficients and $b_{0}$ and $b_\text{crit}$ are the impact parameter evaluated at the closest approach, $r_{0}$, and critical impact parameter, respectively. The first term in Eq. \eqref{eq.48} is the result of the diverging integral and the second term is the result of the regular integral. The coefficients $\bar{a}$ and $\bar{b}$ are expressed as \cite{Tsukamoto:2016jzh},
\begin{equation}
\label{eq.49}
\bar{a} = \sqrt{\frac{2B(r_\text{ps})A(r_\text{ps})}{2A(r_\text{ps}) - A''(r_\text{ps})r_\text{ps}^2}},
\end{equation}
and
\begin{equation}
\label{eq.50}
\bar{b} = \bar{a} \log\left[r_\text{ps} \left( \frac{2}{r_\text{ps}^2}-\frac{A''(r_\text{ps})}{A(r_\text{ps})}\right) \right]+I_{R}(r_\text{ps})-\pi,
\end{equation}
where $A(r_\text{ps})$ is metric function evaluated at the photonsphere, and $I_{R}$ is the regular integral evaluated from 0 to 1. The double prime in Eq. \eqref{eq.49} and Eq. \eqref{eq.50} correspond to the second derivative with respect to $r$ evaluated over $r_\text{ps}$. 

Applying Eq. \eqref{eq.49} to the black hole metric would yield the coefficient $\bar{a}$,  
\begin{equation}
\label{eq.51}
\bar{a} = \frac{r^{p/2}_\text{ps}}{\sqrt{(p^{2}+p-2)M+r^{p}_\text{ps}}},
\end{equation}
where $p = n-3$. When $p=1$, we have the Schwarzschild result of $a=1$. A pattern was observed from $\bar{a}$ which can be simplified as $\bar{a} = \sqrt{p}/p$. As we increase the dimension of the black hole the value of $\bar{a}$ decreases exponentially. The argument of the natural logarithmic term of Eq. \eqref{eq.50} becomes,
\begin{equation}
\label{eq.52}
\bar{b} = \bar{a} \ln\left[2p+4\right]+I_{R}(r_\text{ps})-\pi,
\end{equation}
and we retrieve the Schwarschild result when $p = 1$ (which is 6 as in Ref. \cite{Tsukamoto:2016jzh,Bozza:2002zj}). The regular integral $I_{R}$ is defined as, 
\begin{equation}
\label{eq.53}
I_{R}(r_{0}) \equiv \int^{1}_{0}f_{R}(z, r_{0}) - f_{D}(z,r_{0}) dz,
\end{equation}
where the $f_{R}(z, r_{0})$ was generated from the expansion of the trajectory in Eq. \eqref{eq.43}, which gives us 
\begin{equation}
\label{eq.54}
f_{R}(z, r_{0}) =\frac{2r_{0}}{\sqrt{G(z, r_{0})}}, 
\end{equation}
where $G(z, r_{0}) = RCA(1-z)^{4}$. Notice that $C$ and $A$ are the metric functions for which the position $r$ is expressed in terms of $z$ and $r_{0}$, while $R$ is shown in Eq. \eqref{eq.44}. The generated expression from Eq. \eqref{eq.54} is, 
\begin{equation}
\label{eq.55}
f_{R}(z, r_\text{ps}) =\frac{2r_\text{ps}}{\sqrt{\sum_{m=2}^{m}c_{m}(r_\text{ps})z^{m}}},
\end{equation}
when we evaluate $r_{0}=r_\text{ps}$. On the other hand the $f_{D}(z, r_\text{ps})$ is expressed as,
\begin{equation}
\label{eq.56}
f_{D}(z, r_\text{ps}) = \frac{2r_\text{ps}}{\sqrt{c_{2}z^{2}}},
\end{equation}
where the $c$'s are coefficients of the new variable $z$. For the equations in \eqref{eq.51}, \eqref{eq.52}, and by evaluating the integral in Eq. \eqref{eq.53}, our results are summarized in Table \ref{tab:tab1}.

\begin{table}[!ht]
\centering
    \begin{tabularx}{0.7\textwidth}{>{\centering\arraybackslash}X>{\centering\arraybackslash}X>{\centering\arraybackslash}X>{\centering\arraybackslash}X}
        \hline
        \hline
        \textbf{n} & \bm{$\bar{a}$} & \bm{$2(n-3)+4$}& \bm{$I_{R}(r_\text{ps})$} \\
        \hline
        4(\textbf{Schw}) & $1$ (\textbf{Schw}) & 6 (\textbf{Schw}) & 0.412 (\textbf{Schw})\\
        \hline
        5 & $\sqrt{2}/2$ & 8 & 0.98 \\
        \hline
        6 & $\sqrt{3}/3$ & 10 & 1.03992 \\
        \hline
        7 & $\sqrt{4}/4$ & 12& 1.0986 \\
        \hline
        8 & $\sqrt{5}/5$ & 14& 1.15239 \\
        \hline
    \end{tabularx}
    \caption{The table shows the numerical values of the essential parts of the strong deflection angle. Note that $n$ are the dimensions, $\bar{a}$ is the coefficient, $2(n-3)+4$ is the argument of the logarithmic term, and $I_{\rm R}$ is the regular integral.}
    \label{tab:tab1}
\end{table}

The results in Table \ref{tab:tab1} are consistent with results when using the Schwarzschild metric where $n=4$ as shown in Refs. \cite{Tsukamoto:2016jzh,Bozza:2002zj,Tsukamoto:2021fsz,Chagoya:2020bqz, Tsukamoto:2014dta} with an extension to higher dimension black holes. As the dimensionality increases, the coefficient $\bar{a}$ decreases which significantly affects $\bar{b}$ and leaves the $I_{\rm R}$ to dominate the expression. The increasing value of $I_{\rm R}$ greatly influences the decrease of the deflection angle for higher dimensions.

\begin{figure*}[!ht]
    \centering
    \includegraphics[width=0.6\textwidth]{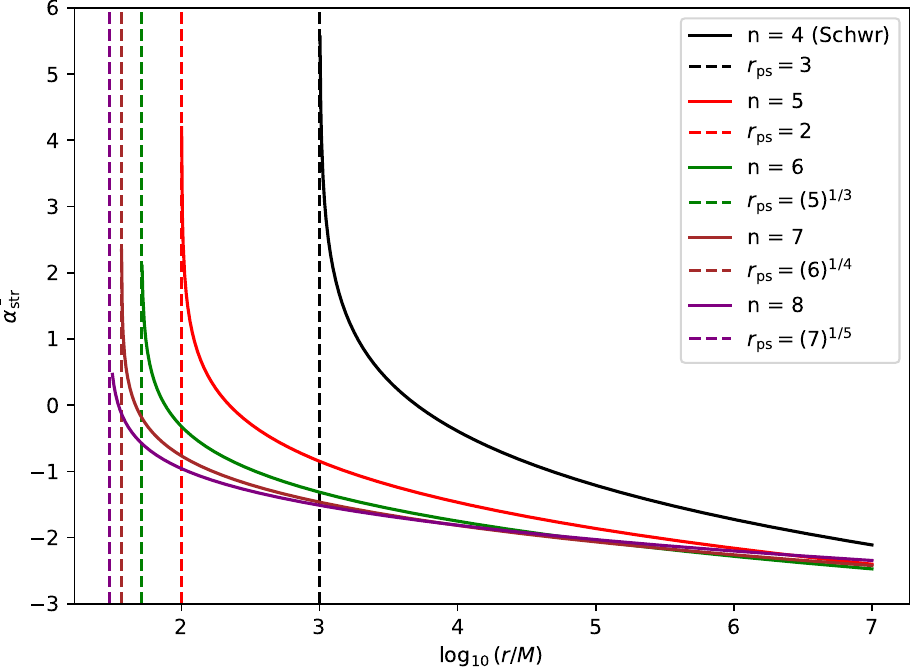}
    \caption{The behavior of $\hat{\alpha}_{\rm str}$ for dimensions $n = 4$ to $n=8$.}
    \label{fig:sda}
\end{figure*}

As shown in Fig. \ref{fig:sda}, the strong deflection angle at higher dimensions exhibits the same behavior but in decreasing values \cite{Belhaj:2020rdb} as dimensionality increases. These decreased values can be attributed to the increase in the number of degrees of freedom in the higher dimensional spaces \cite{Emparan:2008eg}. Investigating the existence of higher dimensions from $n=4$ to $n=7$ is theoretically possible since, at this region, the strong deflection angle is relatively large compared to the weak deflection angle. From $n=8$, the strong deflection becomes significantly small and would require ultrasensitive devices to probe. Employing the methodology outlined in this paper, we successfully computed the higher-dimensional black hole without omitting any dimensions. This contrasts with the approach described in \cite{Tsukamoto:2014dta}, where multiple dimensions were excluded from the calculation. 

%%%%%%%
\section{Quasinormal modes}\label{sec4}
In light of recent observations of gravitational waves \cite{LIGOScientific:2016aoc}, the study of QNMs also becomes an interesting avenue for the study of extra-dimensional models \cite{Chrysostomou:2022evl}. To complement the study of photonspheres above, we now look at how to calculate QNMs in this spacetime with a neural network approach.

The use of neural networks, detailed below, is for its robustness to a range of partial differential equations, as we have here, with a consideration of different dimensionalities. In this section we shall rewrite the metric as \cite{PhysRevD.104.084066}:
\begin{equation}
\label{eq. 57}
    ds^2 = f(r)dt^2 - f^{-1}(r)dr^2 - r^2d\Omega^2_{n - 2},
\end{equation}
where $f(r) = 1 - r^{3 - n}$ when working in the $2M = 1$ units. The equations describing the perturbation of this metric to produce damped sinusoids, the QNMs, are given as~\cite{PhysRevD.104.084066, Konoplya:2003ii}: 
\begin{equation}
\label{eq.58}
    \psi^{\prime\prime} + \left\lbrace \omega^2 - f(r) \left[ \frac{\ell (\ell + n - 3)}{r^2} + \frac{(n - 2)(n - 4)}{4r^2} + \frac{(1 - j^2)(n - 2)^2}{4 r^{n-1}} \right] \right\rbrace \psi = 0, 
\end{equation}
where the prime denotes derivatives with respect to the tortoise co-ordinate $x$. Here $j$ (i.e. spin of perturbing field) is assigned $0, 2, 2/(n - 2)$ for massless scalar, gravitational vector and electromagnetic vector perturbations, respectively. The QNMs, represented here by the perturbation quantities $\psi$, are solutions to the Schr\"{o}dinger-like eigenvalue problem given in Eq. \eqref{eq.58}. and are indexed by the spin $j$, multipole number $\ell$ and overtone number $N$ (capitalised to distinguish it from $n$ denoting spacetime dimensions), with the least-damped mode being $N = 0$. The eigenvalues of Eq. \eqref{eq.58} are the frequencies of the damped sinusoids denoted as $\omega = \omega_{Re} - i\omega_{Im}$, with $\omega_{Re}$ signifying the physical oscillation frequency and $\omega_{Im}$ being associated with the damping rate. These frequencies are important for probing the parameters of the perturbed source. 

\begin{figure*}[!ht]
    \centering
    \includegraphics[width=0.32\textwidth]{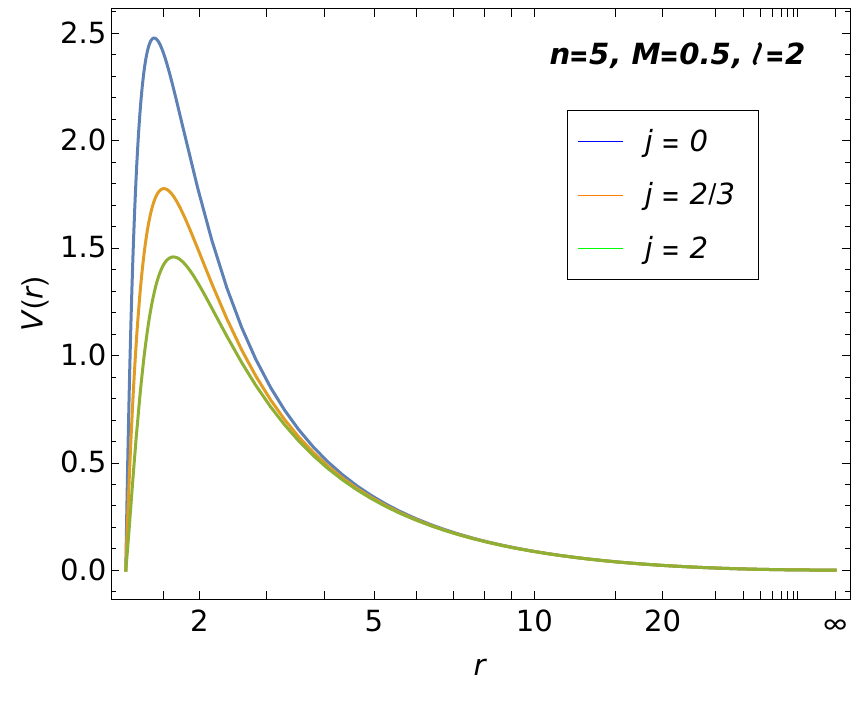}
    \includegraphics[width=0.32\textwidth]{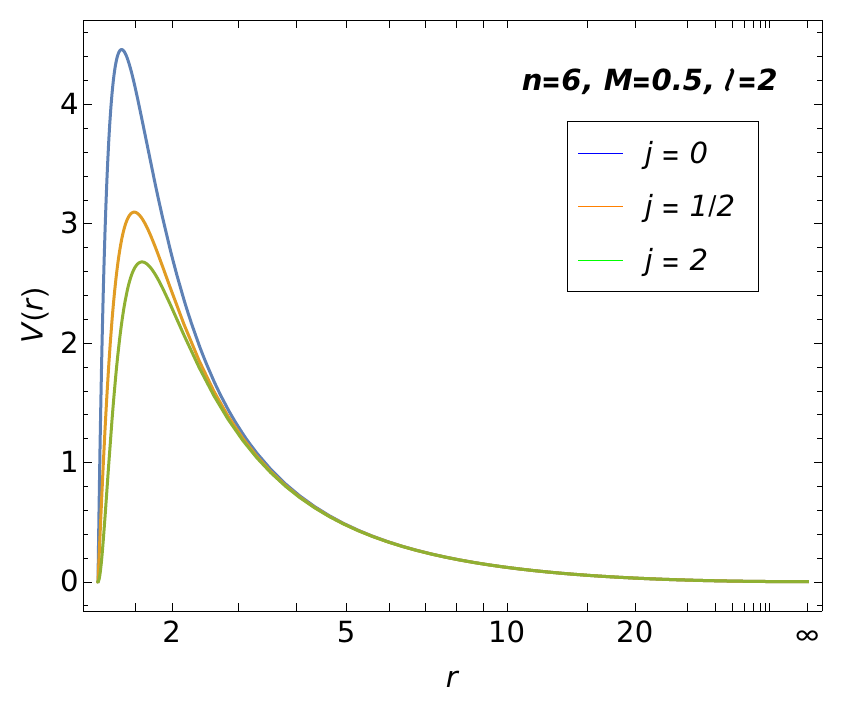}
    \includegraphics[width=0.32\textwidth]{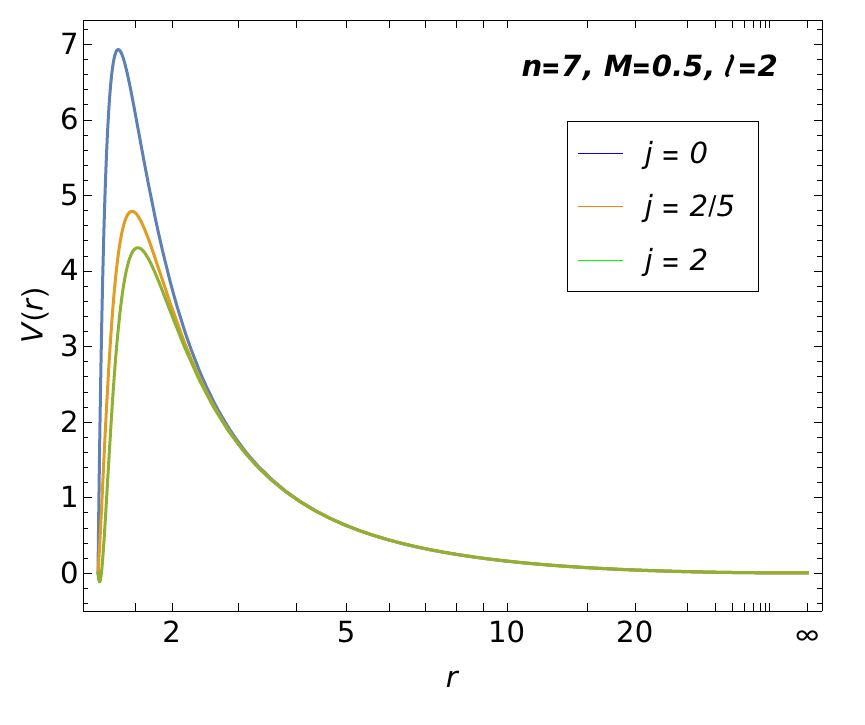}
    \label{effective potential plot}
    \caption{The effective potentials $V(r)$ for perturbations of a Schwarzschild-Tangherlini BH in $n = 5, 6$ and $7$ dimensions. They are ``WKB-well-behaved''; that is, they have a single extremum and they decay at the event horizon $r = 1$ and at spatial infinity.}
\end{figure*}

In the astrophysical setting, the asymptotic behavior of physically allowed modes constitutes only in-going waves to the black hole horizon and only outgoing waves to spatial infinity, in the radial domain of Eq. \eqref{eq.58}, where the radial and tortoise co-ordinates ($r \in [1, \infty)$ and $x \in (-\infty, \infty)$, respectively) are related through the differential equation $dr/dx = f(r)$. Factoring out the asymptotic behavior, $\psi(r)$ can be expressed in the form \cite{PhysRevD.104.084066}:
\begin{equation}
\label{eq.59}
    \psi(r) = 
    \begin{cases}
        \left(\frac{r - 1}{r}\right)^{i \omega/(n - 3)} e^{i \omega r} \chi(r), & \text{if $n$ even,}\\
        \left(\frac{r - 1}{r + 1}\right)^{i \omega/(n - 3)} e^{i \omega r} \chi(r), & \text{if $n$ odd.}\\
    \end{cases}
\end{equation}

Substituting the asymptotic behavior in the perturbation equation and transforming to a finite co-ordinate $\xi = 1 - (1/r)$ or $\xi = 1/r$, both of which yield a finite domain $\xi \in [0, 1]$, the differential equation can be written in the form implicitly incorporating the physical behavior. For example, when working with $\xi = 1/r$ and $n = 5, 6$ and $7$, for which some QNMs are computed using PINNs, the perturbation equations are expressed as:
\begin{align}
\label{eq.60}
 & -((-1 + \xi)(-((1 + \xi)(3 + 4\ell(2 + \ell) - 9(-1 + j^2)\xi^2)) + 6i\xi(1 + \xi)(2 + \xi)\omega + (2 + \xi)^2(3 + \xi)\omega^2)\chi) &\notag\\
 & + 4(-1 + \xi^2)((-2\xi + 4\xi^3 - i(-2 + \xi^2(3 + \xi))\omega)\chi^{\prime} + \xi^2(-1 + \xi^2)\chi^{\prime\prime})) = 0,\ \text{for}\ n = 5, &
\end{align}
\begin{align}
\label{eq.61}
 & (-1 + \xi)(9(1 + x + \xi^2)(2 + \ell(3 + \ell) - 4(-1 + j^2)\xi^3) - 6i\xi(1 + \xi + \xi^2)(1 + 2\xi(3 + \xi))\omega & \notag\\
 & - (6 + \xi(2 + \xi)(3 + \xi))(1 + \xi(4 + \xi))\omega^2)\chi + 3(-1 + \xi^3)(3\xi(-2 + 5\xi^3)\chi^{\prime} - 2i(-3 + \xi^2 + 4\xi^3 + \xi^4)\omega\chi^{\prime} & \notag\\
 & + 3\xi^2(-1 + \xi^3)\chi^{\prime\prime}) = 0,\ \text{for}\ n = 6, &
 \end{align} 
 \begin{align}
 \label{eq.62}
 & ((1 + \xi^2)(15 + 4\ell(4 + \ell) - 25(-1 + j^2)\xi^4) - 4i\xi(1 + 7\xi^2 + 6\xi^4)\omega - (4 + 15\xi^2 + 9\xi^4)\omega^2)\chi &\notag\\
 & + 4(1 + \xi^2)((-2\xi + 6\xi^5 - i(-2 + \xi^2 + 3\xi^4)\omega)\chi^{\prime} + \xi^2(-1 + \xi^4))\chi^{\prime\prime}) = 0,\ \text{for}\ n = 7. &
\end{align}

\newcolumntype{C}[1]{>{\centering\let\newline\\\arraybackslash\hspace{0pt}}m{#1}}

\begin{table}[htp]
\centering
    %\begin{tabularx}{0.7\textwidth}{>{\centering\arraybackslash}X>{\centering\arraybackslash}X>{\centering\arraybackslash}X>{\centering\arraybackslash}X}
    \begin{tabular}{C{0.05\linewidth}C{0.05\linewidth}C{0.25\linewidth}C{0.25\linewidth}C{0.25\linewidth}}
        \hline
        \hline
        \textbf{$j$}     & \textbf{$\ell$}        & \textbf{$n = 5$}     & \textbf{$n = 6$} &  \textbf{$n = 7$} \\[0.5em]
        \hline
\multirow{2}{0.5em}{0}   & \multirow{2}{2em}{0}   & $0.266915041 - 0.191700082i$  & $0.444619726 - 0.266478737i$ & $0.635143574 - 0.332764826i$\\[0.5em]
                         &                        & $(-0.0010\%)(0.0065\%)$       & $(-3.6084\%)(5.9562\%)$      & $(-4.2043\%)(3.8647\%)$\\[0.5em]
                         \cline{2 - 5}
                         & \multirow{2}{2em}{1}   & $0.508001630 - 0.181176678i$  & $0.723261349 - 0.254593735i$ & $0.940660393 - 0.32045074i$\\[0.5em]
                         &                        & $(-0.0013\%)(0.0070\%)$       & $(0.5730\%)(-2.3128\%)$      & $(1.3509\%)(-4.7274\%)$\\[0.5em]
                         \cline{2 - 5}
                         & \multirow{2}{2em}{2}   & $0.755277578 - 0.178759872i$  & $1.005667366 - 0.250980719i$ & $1.248397661 - 0.315932284i$\\[0.5em]
                         &                        & $(-0.0008\%)(-0.0049\%)$      & $(-0.0051\%)(-0.0736\%)$     & $(-0.0121\%)(-0.0749\%)$\\[0.5em]
                         \cline{1 - 5}
\multirow{2}{0.5em}{$\frac{2}{n - 2}$} 
                         & \multirow{2}{2em}{1}   & $0.476377682 - 0.175365997i$  & $0.699958417 - 0.249103188i$ & $0.922894349 - 0.315481801i$\\[0.5em]
                         &                        & $(0.0029\%)(-0.0021\%)$       & $(0.4632\%)(-1.8393\%)$      & $(1.1463\%)(-3.9326\%)$\\[0.5em]
                         \cline{2 - 5}
                         & \multirow{2}{2em}{2}   & $0.734256038 - 0.176224901i$  & $0.989160520 - 0.248210728i$ & $1.234996886 - 0.313165017i$\\[0.5em]
                         &                        &$(-0.0019\%)(0.0069\%)$        & $(-0.0118\%)(0.0261\%)$      & $(-0.0649\%)(0.1520\%)$\\[0.5em]
                         \cline{2 - 5}
                         & \multirow{2}{2em}{3}   & $0.988206324 - 0.176456349i$  & $1.276644568 - 0.247822462i$ & $1.546545329 - 0.312138598i$\\[0.5em]
                         &                        & $(-0.0020\%)(-0.0132\%)$      & $(-0.0052\%)(0.0296\%)$      & $(-0.0102\%)(0.1256\%)$\\[0.5em]
                         \cline{1 - 5}
\multirow{2}{0.5em}{2}   & \multirow{2}{2em}{2}   & $0.566993256 - 0.163759418i$  & $0.762773806 - 0.236783346i$ & $0.967209529 - 0.306153563i$\\[0.5em]
                         &                        & $(-0.0015\%)(-0.0012\%)$      & $(-1.2135\%)(-1.6122\%)$     & $(-1.6652\%)(0.7651\%)$\\[0.5em]
                         \cline{2 - 5}
                         & \multirow{2}{2em}{3}   & $0.862685392 - 0.166923448i$  & $1.093952893 - 0.233947042i$ & $1.318671249 - 0.297263010i$\\[0.5em]
                         &                        & $(0.0004\%)(0.0017\%)$        & $(0.0515\%)(-0.7597\%)$      & $(-0.1828\%)(-1.1075\%)$\\[0.5em]
                         \cline{2 - 5}
                         & \multirow{2}{2em}{4}   & $1.140247654 - 0.170029699i$  & $1.411930008 - 0.236224876i$ & $1.662463146 - 0.297274533i$\\[0.5em]
                         &                        & $(0.0003\%)(0.0124\%)$        & $(0.0227\%)(-0.0997\%)$      & $(0.0211\%)(-0.3870\%)$\\[0.5em]
        \hline
    \end{tabular}
    \caption{PINN approximations of the QNMs frequencies ($\omega_{_{PINN}}$) for $n = 5, 6$ and $7$ Schwarzschild-Tangherlini black holes perturbed by fields of various spins $j$. We compare the PINN approximations with values obtained using separate methods (i.e. the continued fraction method for $n = 5$, see Ref.~\cite{PhysRevD.104.084066}, and the 6-th order WKB method for $n = 6$ and $7$). The values in the parentheses are percentage deviations from comparing with the reference values.}
    \label{tab:tab2}
\end{table}

The corresponding equations in terms of $\xi = 1 - (1/r)$ are similarly second-order, linear, homogeneous differential equations. They are solved here using physics-informed neural networks (PINNs) and the sixth-order WKB method (for comparison between two methods), where the latter is a method that was developed by Ref.~\cite{Konoplya:2003ii}. The applicability of the WKB method in the Schwarzschild-Tangherlini case can be inferred from the related effective potential, which should be a potential barrier with a single extremum point between the event horizon and spatial infinity boundaries where the potential decays. This is seen in Fig \ref{effective potential plot}, for various perturbing fields including gravitational vector perturbations ($j = 2$), for $\ell = 2$ and $n = 5, 6,$ and $7$. The PINN algorithm is a different approximation method; it is machine learning-based and uses highly parameterized ansatzes to approximate the solutions of a differential equation. The approximation is facilitated by an optimization algorithm in which the deviation of the approximate function from the target functions, quantified using the loss function, is minimized by backpropagating the derivatives of the loss function with respect to the ``ansatz'' parameters to take iterative steps towards the loss's global minimum. Let $\hat{\chi}(\xi, \theta)$ be the output value of the neural network (the approximation of the solutions to Eqs. \eqref{eq.60} - \eqref{eq.62}), then $\xi$ is set as an input node to pass into the neural network some data sampled from  $\xi \in [0, 1]$. The parameters to be optimized are denoted by $\theta$. As such, the neural network is represented in general as~\cite{lu2021deepxde}:
\begin{equation}
\label{eq.63} 
\mathcal{N}^{\ell}(\mathbf{\xi}, \mathbf{\theta}) = 
\sigma\left(\mathcal{N}^{\ell- 1}(\mathbf{\xi})\mathbf{W}^{\ell} + \mathbf{b}^{\ell}\right),\quad \text{for $1 \leq \ell \leq L$} , 
\end{equation}
where the neural network output $\mathcal{N}^{\ell = L}(\mathbf{\xi}, \theta) = \hat{\chi}(\xi, \theta)$ is a composite function of recursive linear transformations of the hidden layers $\mathcal{N}^{L - 1}, \mathcal{N}^{L - 2}, ... \mathcal{N}^{1}$ as given within the parentheses of the right-hand side of Eq. \eqref{eq.61}. Here $\mathbf{W}^{\ell}$ and $\mathbf{b}^{\ell}$ are the weight matrices and bias vectors, respectively, which we represent collectively with the term $\theta$. To enhance function approximation, an activation function $\sigma$ is applied to each hidden layer $\ell \in [1, L - 1]$.   
We use PINNs here to approximate $\omega$ for $n= 5, 6$ and $7$ Schwarzschild-Tangherlini black holes using a basic deep neural network set-up consisting of two fully connected layers and the adaptive moment estimation (ADAM) algorithm as the optimizer. Table~\ref{tab:tab2} lists the approximate QNM frequencies generated by PINNs whose loss function we actively tuned toward $N = 0$ QNMs. The loss function minimized by the ADAM optimizer is given as:
\begin{equation}
    \mathcal{L} = \langle[\mathcal{D}(\hat{\chi})]^2\rangle + \frac{1}{\langle\hat{\chi}^2\rangle} + \frac{1}{\langle\hat{\omega}^k_{Re}\rangle},
\end{equation}
which consists of the differential equation residual $\mathcal{D}(\chi)$, where $\mathcal{D}$ is the differential operator in Eqs. \eqref{eq.60}-\eqref{eq.62}. The extra two terms constrain the space of eigenvalues, to be learned during optimisation, to within the fundamental mode values, which facilitates convergence to a single solution.  Note that for the $n = 6$ and $7$, for various multipole numbers, the determination of the optimal choice of change of variable (i.e. $\xi = 1/r$ or $\xi = 1 - (1/r)$) and choosing between $k = 0, 2, 4$ in the third term is a heuristic process and there is no one optimal set of parameters. We compared the $n = 5$ QNMs with Ref.~\cite{PhysRevD.104.084066} to estimate the accuracy of PINNs, given the high level of accuracy of the continued fraction method used in Ref.~\cite{PhysRevD.104.084066}. As shown in the $n = 5$ case in Table~\ref{tab:tab2}, the PINN approximations generally differ from the CFM by $\ll0.1\%$. When comparing with the sixth-order WKB method for $n = 6$ and $7$ the percentages in the parentheses quantify the consistency of the QNMs between the WKB method and PINNs, where neither method is assumed the more accurate than the other. As can be seen, the QNMs agree more with increasing $\ell$, which is when the WKB method is expected to improve in accuracy. It could be deduced from this behaviour that we obtain higher accuracy computations of QNMs from PINNs than the sixth-order WKB method for lower multipole numbers.
\begin{figure}[ht!]
\begin{center}
\includegraphics[width=0.48\textwidth]{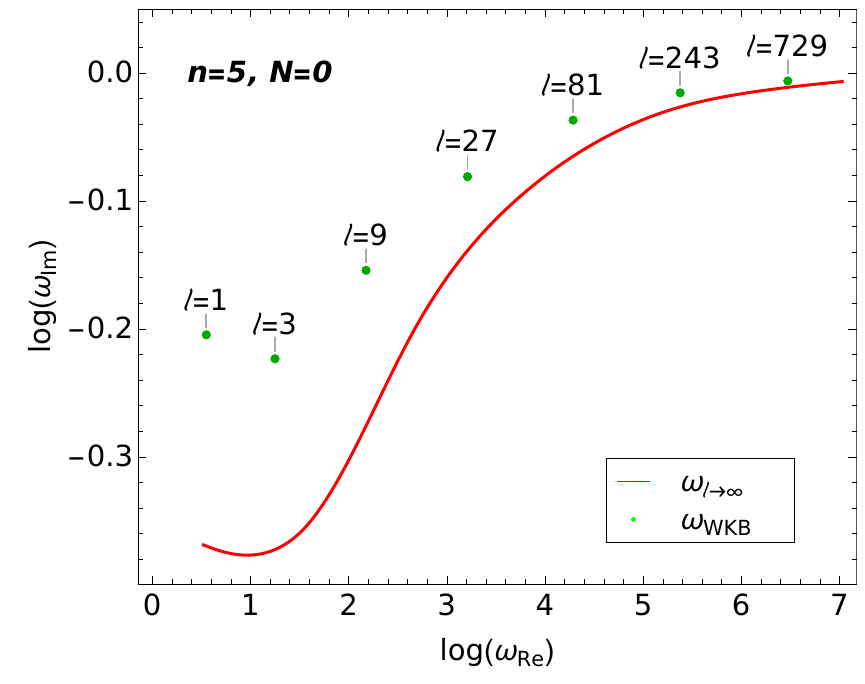}
\includegraphics[width=0.48\textwidth]{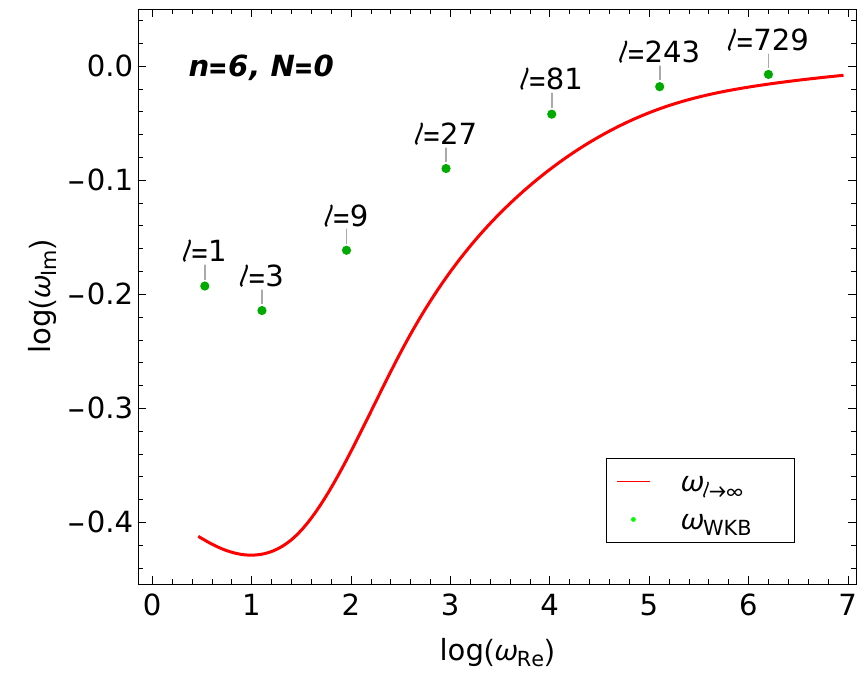}
~
\includegraphics[width=0.48\textwidth]{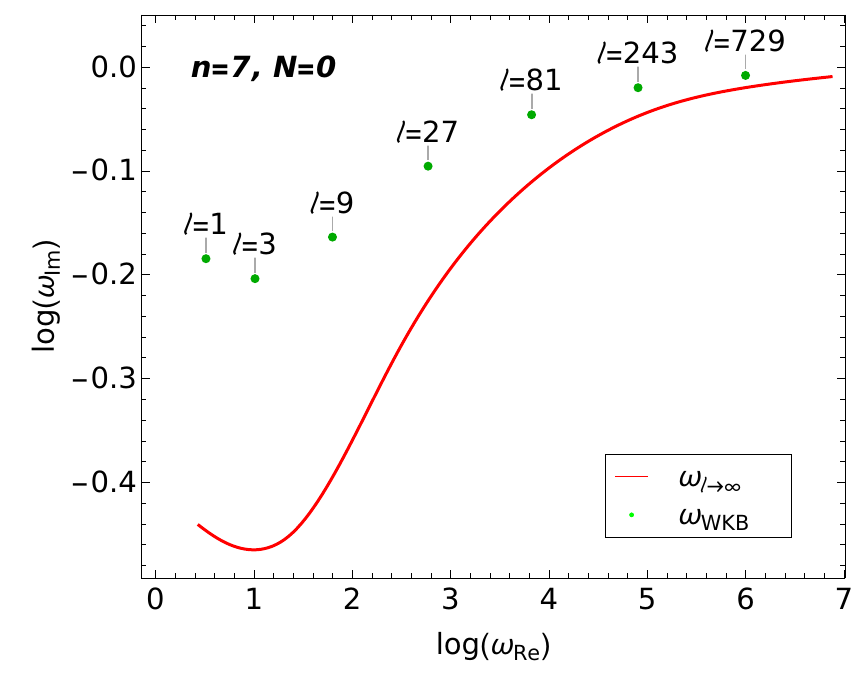}
\includegraphics[width=0.48\textwidth]{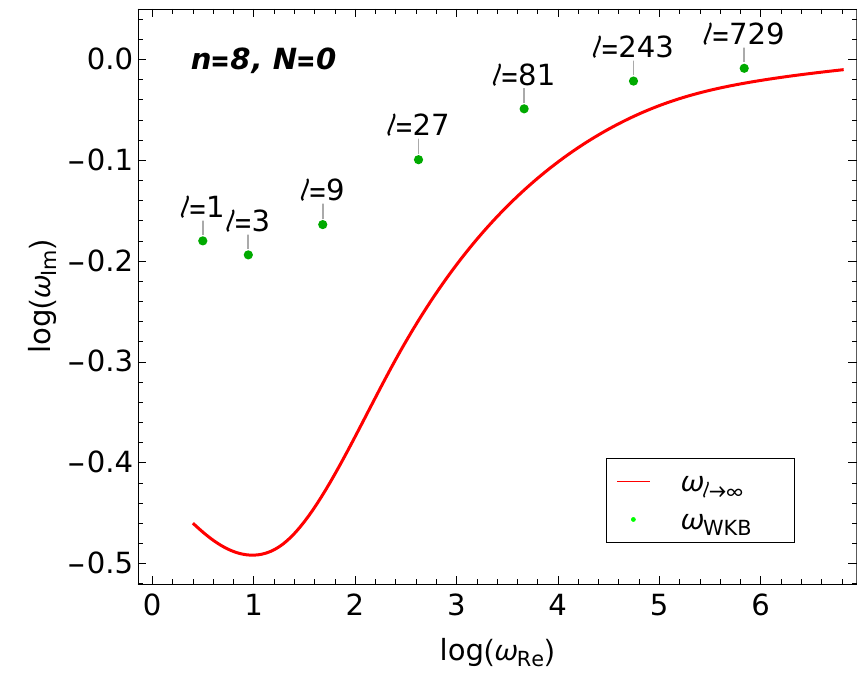}
\end{center}
\caption{\label{eikonal-limit}The approach of the fundamental mode and spin $2/(n - 2)$ QNMs towards the eikonal limit relation $\omega_{\ell \rightarrow \infty} =\Omega_c\ell - i(N + 1/2)|\lambda|$ (represented by the red curve).
}
\end{figure}

We are interested in the correspondence between the dynamics of the photon sphere of a Schwarzschild-Tangherlini black hole and their QNMs produced in the wake of perturbations by various test fields. The correspondence was determined by Ref. \cite{Cardoso:2008bp, KONOPLYA2023137674} to be:  
\begin{equation}
\label{eq. 64}
    \omega_{N} = \Omega_c\ell - i\left(N + \frac{1}{2}\right)|\lambda|,\ \ell \ll N,
\end{equation}
where $\Omega_c$ and $\lambda$ are the orbital frequency and the Lyapunov exponent (i.e. the instability time scale), respectively, at the black hole's photon sphere. Note that Eqn~\eqref{eq. 64} does not apply in all QNM computations, such as when considering gravitational perturbations of arbitrary asymptotically de Sitter BHs \cite{KONOPLYA2023137674, KONOPLYA2017597}. However, we can still invoke the correspondence in the asymptotically flat Schwarzschild-Tangherlini case, which is demonstrated in Fig \ref{eikonal-limit}. For varying $\ell$ and $n$, QNMs plotted in Fig \ref{eikonal-limit} are obtained using the WKB method, where in the eikonal limit the tendency towards the properties of the photonsphere (i.e. Eqn. \eqref{eq. 64}) is seen. Physically, in such cases, massless perturbing fields that have sufficient energy can orbit near the black hole in the location of the unstable circular geodesic from where they are either drawn inwards or escape outward to spatial infinity. As such the attenuation of QNMs in the temporal domain, quantified by $\omega_{Im}$, is explained by the leaking of particles from the unstable orbits in a time scale given by the Lyapunov exponent.

Additionally, the effect of dimensionality can be seen in QNMs as is done with the strong deflection angle, for example. For an increase in the number of dimensions, the imaginary part of $\omega$ is higher for comparable physical oscillation frequencies across different values of $n$ (see Table \ref{tab:tab2}). In other words, there is an increased rate of leaking of particles (quantified by the Lyapunov exponent) at the unstable orbits, for higher dimensions. Regarding $\omega_{Re}$, Konoplya \cite{Konoplya:2003ii} showed that, for massless scalar fields and a given $\ell$, the values lie on strict line; that is, they are directly proportional to $r_0 n$, where $r_0 = 2M$ determines the units of $\omega$. Here, we see similar behavior across the different perturbing fields considered (whose QNMs are in Table \ref{tab:tab2}) with $\omega_{Re}$ increasing with $kn$, with $k$ as some constant of proportionality. 

%%%%%%%
\section{Conclusion}\label{conc}
In this paper, we have studied the effect of the number of dimensions $ n $ on the black hole’s photonsphere, shadow, deflection angle (both in the weak and strong regimes), and quasinormal modes. First, we have seen how the photonsphere $ r_{ps} $ decreases with increasing $ n $. Results indicate that the rate at which $ r_{ps} $ changes relative to $ n $ becomes smaller as $ n $ increases. As the shadow radius $ R_{sh} $ is related to $ r_{ps} $, we have seen the same behavior. For instance, increasing $ n $ results in decreased $ R_{sh} $.

We then analyzed how the deflection angle changes with $ n $ by deriving the most general formula for the weak deflection angle $ \alpha $; not only accommodating the finite distance of the source and observers but also valid for both time-like and null particles. As higher dimensionality is included, the formula necessitates the existence of a hypergeometric function, making the calculation of $ \alpha $ more precise. If $ \alpha $ is due to the time-like particles, we have seen how it deviates from the null result, depending on its speed $ v $. Finally, the results imply that probing the existence of higher dimensions using weak deflection angle (as well as the lens images that it can cause) requires ultra-sensitivity as increased $ n $ results in small $ \alpha $. We also examined the strong deflection angle that occurs near the $ r_{ps} $. Dimensionality and gravitational lensing are correlated, as can be inferred from the plot in Fig. \ref{fig:sda} that the strong deflection angle around a higher dimensional black hole decreases as the dimension, $ n $, increases. The degrees of freedom rise in proportion to dimensionality, which impacts the gravitational interaction surrounding the black hole. Such a decrease in the strong deflection angle raises the possibility that the gravitational lensing phenomenon is dampened due to higher dimensions leading to more difficulty of measurement. In addition, the methodology employed in this paper generalizes the approach in \cite{Tsukamoto:2014dta}, which opens the avenue to calculate the strong deflection in all dimensions. Investigating the precise mechanisms underlying this relationship may yield important new information about how gravity behaves in higher-dimensional regions and how this may affect astrophysical occurrences. Since the strong deflection exhibits large values as it approaches the photon sphere, higher dimensions can possibly be detected when investigated in these regions. 

Given the increasing difficulty in measuring gravitational phenomena as the number of dimensions increases, QNMs in higher dimensions offer another means to probe the photonsphere and related properties about the effect of Schwarzschild-Tangherlini black holes on interacting fields. This is made possible by the known relation between QNMs and the dynamics of the photonsphere in the eikonal limit, which are described by the orbital frequency and Lyapunov exponent. With the decrease of the black hole photonsphere, shadow, and deflection angle, there is an increase in both the physical oscillation frequency and damping time-scale of QNMs (as shown Table \ref{tab:tab2}). Knowing the correspondence between QNMs and the parameters of the photonsphere in the eikonal limit, QNMs could offer a useful indirect probe of the optical properties of the black hole which may not be accessible through direct measurement. This could be especially significant as observational technologies evolve, potentially enabling detection of higher-dimensional effects that are currently beyond our reach.

As a final remark, the uncertainties in the Event Horizon Telescope’s measurement of the shadow radius at $3\sigma$ level for Sgr. A*, and $1\sigma$ level for M87* \cite{EventHorizonTelescope:2019dse,EventHorizonTelescope:2022wkp,EventHorizonTelescope:2021dqv,Vagnozzi:2022moj}, suggest that only $n=4$ is included by observation, unless one may speculate with fractal dimensions \cite{Calcagni:2017ymp} and once a $3\sigma$ confidence level becomes available for M87*. Currently, these findings imply that dimensions $n>5$ are not currently supported by current astrophysical observations and technologies. While our findings also primarily support $ n=4 $, consistent with current observations, theoretical exploration of higher dimensions remains essential. These studies advance quantum gravity models and deepen our understanding of black hole behavior, particularly in regimes where direct astrophysical evidence is presently unavailable.

Further research includes the possibility of extending this study to incorporate black hole charge $Q$, or, in a more general sense, the spin parameter $a$, would allow for greater comparison with physically observable phenomena. While the Schwarzschild-Tangherlini black hole only specializes within whole number integers of $n$ dimensions, it is also important to explore black holes in multi-fractional theory \cite{Calcagni:2017ymp} due to its applicability in the quantum realm, even allowing dimensions higher than $n = 4$. Lastly, while this manuscript focuses on a simplified treatment to isolate the effects of dimensionality \( n \), future studies could explore compactification schemes or brane-world frameworks to examine their impact on black hole physics. Such approaches could offer more realistic models aligned with quantum gravity theories, extending the baseline understanding provided here of higher-dimensional effects on black hole properties.

%%%%%%%
\acknowledgments
R. P. would like to acknowledge networking support of the COST Action CA18108 - Quantum gravity phenomenology in the multi-messenger approach (QG-MM), COST Action CA21106 - COSMIC WISPers in the Dark Universe: Theory, astrophysics and experiments (CosmicWISPers), the COST Action CA22113 - Fundamental challenges in theoretical physics (THEORY-CHALLENGES), and the COST Action CA21136 - Addressing observational tensions in cosmology with systematics and fundamental physics (CosmoVerse). A. S. C. is supported in part by the National Research Foundation (NRF) of South Africa. A. N. is supported by an SA-CERN Excellence Bursary through iThemba LABS.

\bibliography{ref}
\bibliographystyle{apsrev}
\end{document}